\documentclass[aps,pra,a4paper,superscriptaddress,longbibliography,reprint]{revtex4-1}

\usepackage[T1]{fontenc}
\usepackage{newtxmath,newtxtext}

\usepackage{amsmath}
\usepackage{amsfonts}
\usepackage{bm}
\usepackage{bbm}
\usepackage{subfigure}
\usepackage{siunitx}

\usepackage{graphicx}
\usepackage{ulem}
\usepackage{color}

\newcommand{\eq}[1]{Eq.~(\ref{#1})}
\newcommand{\fig}[1]{Fig.~\ref{#1}}
\newcommand{\etal}{\textit{et al.}}
\newcommand{\ofr}{(\textbf{r})}
\newcommand{\dr}{\text{d}\textbf{r}}
\newcommand{\csixty}{C$_{60}$}

\begin{document}

\title{Machine learning force fields based on local parametrization of dispersion interactions:
Application to the phase diagram of \csixty}

\author{Heikki Muhli}
\affiliation{Department of Applied Physics,
Aalto University, 02150, Espoo, Finland}

\author{Xi Chen}
\email{xi.6.chen@aalto.fi}
\affiliation{Department of Applied Physics,
Aalto University, 02150, Espoo, Finland}

\author{Albert P. Bart\'ok}
\affiliation{Department of Physics and Warwick Centre for Predictive Modelling,
School of Engineering, University of Warwick, Coventry CV4 7AL, United Kingdom}

\author{Patricia Hern\'andez-Le\'on}
\affiliation{Department of Electrical Engineering and Automation,
Aalto University, 02150, Espoo, Finland}

\author{G\'abor Cs\'anyi}
\affiliation{Engineering Laboratory, University of Cambridge,
Cambridge CB2 1PZ, United Kingdom}

\author{Tapio Ala-Nissila}
\affiliation{Department of Applied Physics, QTF Center of Excellence,
Aalto University, 02150, Espoo, Finland}
\affiliation{Interdisciplinary Centre for Mathematical Modelling
and Department of Mathematical Sciences, Loughborough University,
Loughborough, Leicestershire LE11 3TU, United Kingdom}

\author{Miguel A. Caro}
\email{mcaroba@gmail.com}
\affiliation{Department of Electrical Engineering and Automation,
Aalto University, 02150, Espoo, Finland}

\date{\today}

\begin{abstract}
We present a comprehensive methodology to enable addition of van der Waals (vdW) corrections to
machine learning (ML) atomistic force fields. Using a Gaussian approximation potential (GAP)
[Bart\'ok \etal., Phys. Rev. Lett. \textbf{104}, 136403 (2010)] as baseline, we accurately
machine learn a local model of atomic polarizabilities based on Hirshfeld volume partitioning
of the charge density [Tkatchenko and Scheffler, Phys. Rev. Lett. \textbf{102}, 073005
(2009)]. These environment-dependent polarizabilities are then used to parametrize a screened
London-dispersion approximation to the vdW interactions. Our ML vdW model
only needs to learn the charge density partitioning implicitly, by learning the reference Hirshfeld volumes from density
functional theory (DFT). In practice, we can predict accurate Hirshfeld volumes from the knowledge
of the local atomic environment (atomic positions) alone, making the model highly computationally efficient. For additional efficiency,
our ML model of atomic polarizabilities reuses the same many-body atomic descriptors used for the
underlying GAP learning of bonded interatomic interactions. We also show how the method
enables straightforward computation of gradients of the observables, even when these remain
challenging for the reference method (e.g., calculating gradients of the Hirshfeld volumes in DFT).
Finally, we demonstrate the
approach by studying the phase diagram of \csixty{}, where vdW effects are
important. The need for a highly accurate vdW-inclusive reactive force field is
highlighted by modeling the decomposition of the \csixty{} molecules taking place at high
pressures and temperatures.
\end{abstract}

\maketitle

\section{Introduction}

Interatomic interactions that emanate from the underlying quantum-mechanical nature
of matter can often be broken down into effective contributions in terms of classical force fields.
Traditionally, these contributions are four: covalent (or ``bonded''), electrostatics,
Pauli repulsion, and dispersion (or ``van der Waals'').
In the context of classical molecular dynamics (MD) and the development of
atomistic force fields,
covalent interactions typically account for the portion of the potential
energy surface (PES) of a system of interacting atoms that can be parametrized
into a set of harmonic-like short-range functions.
Electrostatic interactions, in their simplest implementation, take the
form of point charges located at the positions of the atomic nuclei, interacting
via a long-range Coulomb potential; more complicated forms can account for atom-centered
multipole expansions and polarizabilities. Repulsion (often termed ``exchange''
or ``Pauli'' repulsion) is the very strong interaction preventing two
atoms from coming too close to one another, a phenomenon that can be ascribed
to Pauli's exclusion principle, and decays very rapidly with interatomic
separation. Finally, ``dispersion'' is an umbrella term of sorts for ``the rest'',
i.e., the difference between the quantum-mechanical energies and forces
and the sum of the covalent, electrostatic and repulsion energies and
forces~\cite{martin_2004}.

Dispersion interactions, also often referred to as ``van der Waals'' (vdW) interactions
(and making up part of the ``correlation'' energy, as referred to
within the electronic structure community), are long ranged,
but typically \textit{individually} weak. E.g., the dispersion energy between
two isolated atoms or molecules is usually much smaller than their corresponding
electrostatic interaction. However, while electrostatic interactions may cancel out
due to the balance between attractive and repulsive energy contributions, dispersion
interactions are typically attractive. Thus, the dispersion interaction can be the driving
force behind interesting emerging physical phenomena, such as the bonding of
2D layers of material to form 3D solids (graphite or black phosphorus being prime
examples). Trending topics in nanoscience and nanotechnology include
``van der Waals heterostructures''~\cite{geim_2013} and, more generally,
simply the desire to
describe atomistic systems to a greater level of detail.

Machine learning (ML) in molecular and materials
modeling~\cite{bartok_2017,deringer_2019,schutt_2020} is an emerging
interdisciplinary field that touches on physics, chemistry, materials science
and biochemistry, and has seen exponential growth in recent years. Still, dispersion
interactions remain elusive. This is particularly true for materials modeling,
where the greater atomic density complicates the description of the system. At the
root of the issue is the need for local atomic descriptors as input for ML
potentials to keep the atomistic problem
tractable~\cite{bartok_2013,himanen_2020}. In ML atomistic
modeling, the representation of the whole system is carried out in practice
by representing the individual building blocks and then adding them
together~\cite{behler_2007,bartok_2010}. One
of these individual blocks can be for example an atom and all of its neighbors
within a given cutoff sphere of a few {\aa}ngstr\"oms. This poses a serious problem for ML
potentials because the effective range of vdW interactions for the
system at hand can be in excess of a few \textit{tens} of {\aa}ngstr\"oms.

To circumvent this issue, a promising solution is to rely on simple analytical
forms for the long-range dispersion interactions between a pair of atoms, which
are computationally cheap to evaluate, and condense all the complicated
many-body physics into the \textit{local parametrization} of said function. This
strategy is in the spirit of existing  popular ``dispersion corrections'' to
density functional theory (DFT)~\cite{grimme_2006}. One of these approaches,
the Tkatchenko-Scheffler (TS) vdW correction~\cite{tkatchenko_2009}, is
precisely the starting point for our present
approach. We will show in this paper that TS corrections, which rely on
computationally expensive integration of the charge density field, can be effectively
and accurately machine learned. Furthermore, we will show how properties that are
not straightforward to compute within the context of the reference method (e.g.,
the gradients of these charge integrals) can be readily and inexpensively obtained
within our ML framework.

Previous recent attempts at machine learning vdW corrections have explored
the ideas of mixing parametric and non-parametric fits~\cite{bereau_2018},
learning highly accurate dispersion energies to add them directly on top of
DFT~\cite{mezei_2020}, or applying preexisting (unmodified) vdW
correction schemes on top of ML potentials~\cite{morawietz_2016}.
The most similar in spirit
to our current approach is the work by Bereau \textit{et al.}~\cite{bereau_2018}, who
developed accurate ``all-ML'' interatomic potentials for small systems, and used a local
parametrization of a physical vdW model [many-body dispersion (MBD)~\cite{tkatchenko_2012}],
which goes beyond the TS model used here. Some of these models achieve highly accurate
results but still fall short of ``force field'' computational efficiency, and are
therefore not amenable to large-scale MD simulations. Therefore, the focus in this paper
is on adding vdW corrections within a general and flexible framework, which can be
easily extended in the future to accommodate more sophisticated vdW models and other
interactions (such as electrostatics), while maintaining a very low computational
footprint. This allows us to perform large-scale MD simulations, both with regards
to the number of atoms and accessible time scales, albeit necessarily sacrificing some
accuracy in the process.

The outline of this paper is as follows. We first introduce the general methodology
for the addition of pairwise dispersion corrections on top of DFT energies. This
is followed by a general discussion of the ML methods we use to construct force
fields, namely GAPs, kernels and many-body atomic descriptors. Then
we discuss how this methodology is applied to learning the dispersion energies
from local parameters of the atomic environments and how we calculate the
dispersion forces analytically. Before moving on to the simulation results, we
also discuss the implementation of our methodology in the TurboGAP and QUIP codes.
We then present a new GAP force field for carbon with vdW corrections, specifically
tailored for simulation of \csixty{} based
systems, and show basic performance and accuracy tests. Finally, with our new
GAP, we characterize the phase stability and phase transformations taking place
in the \csixty{} molecular system over a wide range of pressures and temperatures.

\section{Methodology}

\subsection{Tkatchenko-Scheffler vdW method}

The TS method relies on two fundamental
approximations, one regarding the functional form of dispersion interactions
and another regarding the parametrization of said functional
form~\cite{tkatchenko_2009,hermann_2017,stohr_2019}. First,
there is a range separation of the electron correlation energy, where
the short-range correlation is captured by the underlying DFT functional
(usually, but not necessarily, PBE~\cite{perdew_1996}) and the long-range
correlation is modeled via the London dispersion formula for pairwise
interactions with polynomial decay as $\propto 1/r^6$~\cite{hermann_2017}.
The transition between
the short and long ranges is modeled via a ``damping'' function, an approach
first introduced by Grimme~\cite{grimme_2006}. Second, the novelty in TS
resides in how the London dispersion formula and the damping function are
parametrized, taking the atomic environments of the two interacting atoms
into account. TS assumes an ``atom in a molecule'' approach, which approximates
the properties of the atoms in a molecule or solid as proportional to the
properties of the ``free'' (neutral) atom. The resulting
equations are remarkably simple:
\begin{align}
& E_\text{TS} = - \frac{1}{2} \sum\limits_{i} \sum\limits_{j \neq i} C_{6,ij} \frac{f_\text{damp}
(r_{ij}; r_{ij}^\text{vdW}, s_{\rm R}, d)}{r_{ij}^6},
\nonumber \\
& C_{6,ij} = \frac{2 C_{6,ii} C_{6,jj}}{\frac{\alpha_{j,0}}{\alpha_{i,0}}C_{6,ii} +
\frac{\alpha_{i,0}}{\alpha_{j,0}}C_{6,jj}},
\quad
C_{6,ii} = {\nu_i}^2 C_{6,ii}^\text{free},
\nonumber \\
& \alpha_{i,0} = \nu_i \alpha_{i,0}^\text{free},
\quad
r_{ij}^\text{vdW} = r_{i}^\text{vdW} + r_{j}^\text{vdW},
\quad
r_{i}^\text{vdW} = {\nu_i}^\frac{1}{3} r_{i,\text{free}}^\text{vdW},
\nonumber \\
& \nu_i = \frac{V_i}{V_i^\text{free}} = \frac{\int \dr \, w_i \ofr \rho \ofr}{\int
\dr \, \rho_i^\text{free} \ofr}, \qquad w_i \ofr = \frac{\rho_i^\text{free} \ofr}
{\sum\limits_j \rho_j^\text{free} \ofr},
\nonumber \\
& f_\text{damp}(r_{ij}; r_{ij}^\text{vdW}) = \frac{1}{1 + \exp \left[ - d \left(
\frac{r_{ij}}{s_{\rm R} r_{ij}^\text{vdW}} - 1 \right) \right]}.
\label{TS}
\end{align}
Here, the local environment dependence of the force constants $C_{6,ij}$,
van der Waals radii $r^\text{vdW}_{ij}$
and atomic polarizabilities $\alpha_0$ is taken into account via the (effective) Hirshfeld
volumes $\nu_i$, given as a functional of the electron density (computed at the
DFT level). The various parameters labeled as ``free'' refer to the reference
quantities of the free (isolated and neutral) atoms, and are
tabulated~\cite{grimme_2006,chu_2004}. The damping function used to screen the
TS force field also relies on two empirical parameters, $d=20$ and $s_{\rm R}=0.94$ (for PBE;
$s_{\rm R}$ is functional dependent~\cite{tkatchenko_2009,marom_2011,caro_2017c}). All other
quantities within the TS framework are extracted directly from the electron density
via the Hirshfeld volumes.

For actual simulations to be manageable, a vdW cutoff radius needs to be
introduced at a distance where vdW interactions are expected to be negligible, to make
the sums over atom pairs finite. The cost
of computing pairwise vdW interactions grows approximately as the cube of this cutoff radius.
Therefore, in addition to the TS damping function that screens the dispersion interaction at
short distances, we also introduce a smoothing function that makes the dispersion energies
and their derivatives, that is, the dispersion forces, smooth and continuous at the vdW cutoff
radius, and also takes care of the singularity at $r_{ij} \rightarrow 0$. For computational
reasons, we employ a polynomial function for this purpose:
\begin{align}
E_\text{TS} = - \frac{1}{2} \sum\limits_{i} \sum\limits_{j \neq i} C_{6,ij} \frac{f_\text{damp}
(r_{ij})}{r_{ij}^6} f_{\text{cut}} (r_{ij}; r_{\text{cut}},d_{\text{b}}),
\nonumber \\
f_{\text{cut}} (r_{ij}) = \begin{cases}
 1, & \text{if } r_{\text{c,i}} + d_{\text{b,i}} < r_{ij} \leq r_{\text{c,o}} - d_{\text{b,o}} \\
 1-3r_{\text{b,o}}^2+2r_{\text{b,o}}^3, & \text{if } r_{\text{c,o}} - d_{\text{b,o}} < r_{ij} \leq r_{\text{c,o}} \\
 3r_{\text{b,i}}^3 - 2r_{\text{b,i}}^3, & \text{if } r_{\text{c,i}} < r_{ij} \leq r_{\text{c,i}}+d_{\text{b,i}} \\
 0, & \text{otherwise},
\end{cases} \nonumber \\
r_{\text{b,o}} = \frac{r_{ij}- r_{\text{c,o}}+d_{\text{b,o}}}{d_{\text{b,o}}}, \quad r_{\text{b,i}} = \frac{r_{ij}-r_{\text{c,i}}}{d_{\text{b,i}}},
\label{smooth}
\end{align}
where $r_{\text{c,o}}$ is the cutoff radius for the vdW interaction and $d_{\text{b,o}}$
is the width of the ``buffer'' region where the cutoff function smoothly switches from
$1$ to $0$ as it approaches $r_{\text{c,o}}$.
The same approach, in this case switching from 0 to 1, is used for an inner cutoff $r_\text{c,i}$
at small interatomic separations (usually with an inner buffer region $d_\text{b,i}$
between 0.5 and 1~\AA{}),
to avoid the singularity in the original TS expression for the damping function as
$r_{ij} \rightarrow 0$. The combined effect of the damping function and cutoff functions
can be visualized in the Supplemental Material (SM)~\cite{sm}.

A clear limitation of the TS method is that it neglects many-body
effects~\cite{tkatchenko_2012}, which can be important in particular for accurately
estimating the $C_6$ coefficients (``electronic'' many-body effects)~\cite{otero_2020}.
The D3 method~\cite{grimme_2010}, which also omits many-body
effects, can yield more accurate dispersion energies than TS in some cases and would
be similarly computationally cheap. Our choice of TS as dispersion correction scheme
for this work serves two purposes. On the one hand, it demonstrates
that vdW corrections can be efficiently coupled to ML potentials and used in large-scale MD
simulation. On the other, it paves the way for subsequent improvements that directly
feed on effective Hirshfeld volumes (such as MBD) or can benefit from local parametrization
of a long-range interaction with simple functional form, the most relevant of which would
be long-range electrostatics.

\subsection{Gaussian approximation potentials}

Within the context of the Gaussian approximation potential (GAP)
framework~\cite{bartok_2010,bartok_2015}, which is an ML approach for atomistic modeling
based on kernel ridge regression (KRR), any physical property $f$ that can be assigned
to a local atomic environment, such as a local energy, a force or an effective
Hirshfeld volume, can be written as a linear combination of kernel functions centered on
the training points:
\begin{equation}
    \bar{f}_* (\mathbf{d}_*) = \delta^2 \sum\limits_{t} \alpha_t k(\mathbf{d}_*,\mathbf{d}_t),
    \label{kernel_eq}
\end{equation}
where $\mathbf{d}_*$ is the local atomic descriptor for which the prediction is made,
$\mathbf{d}_t$ are the descriptors of the configurations in the training set,
$k(\mathbf{d}_*,\mathbf{d}_t)$ is a kernel, or similarity measure (bounded between 0
and 1) between the atomic environments characterized by $\mathbf{d}_*$ and $\mathbf{d}_t$,
$\alpha_t$ are a set of fitting coefficients, and $\delta$ is a scaling parameter with
units of the magnitude under study (e.g., for energies $\delta$ is given in eV).
Due to the large amount of available training data points, it is usually beneficial to
sparsify the model such that many fewer fitting coefficients $\alpha_t$ in \eq{kernel_eq} are
used for prediction than there are local environments in the training database (since
the cost of evaluating the model scales linearly in the
size of the sparse set)~\cite{bartok_2015}. The atomic environments present in the training
database are also often repetitious, which means that they do not contribute much new
information to the model compared to the computational cost of including them.

More generally, a GAP can be constructed by addition of several terms of the form of
\eq{kernel_eq}, typically (but not limited to), two-body (2b), three-body (3b) and
many body (mb). A more detailed account of GAP construction, including discussion of
descriptors, sparsification, regularization and training, has been given
elsewhere~\cite{bartok_2010,bartok_2013,bartok_2015,deringer_2017,caro_2019}. For
conciseness, here we restrict ourselves to giving a brief overview of the mb descriptors
used for learning effective Hirshfeld volumes in the context of the present methodology,
and refer the reader to the literature for further details.

The numerical fingerprint of an atomic environment centered on a particular atom
can be obtained from the smooth overlap of atomic positions (SOAP)~\cite{bartok_2013}.
For the SOAP mb descriptor, we define the kernel directly as proportional to
the overlap integral of two atomic density fields $\rho(\mathbf{r})$ and
$\rho'(\mathbf{r})$~\cite{bartok_2013,caro_2019} averaged over all possible relative
orientations $\hat{R}$ of the two atomic environments:
\begin{equation}
    k(\rho,\rho') = \int \mathrm{d}\hat{R} \left| \int \mathrm{d}\mathbf{r} \,
    \rho(\mathbf{r})\rho'(\hat{R}\mathbf{r}) \right|^n,
    \label{SOAP_kernel}
\end{equation}
where the atomic density field is constructed from the sum of individual atomic
contributions within the atomic neighborhood (as defined by a cutoff distance)
of the central atom $i$, possibly including the central atom itself:
\begin{equation}
    \rho(\textbf{r} - \textbf{r}_i; i) = \sum_{j} g( \textbf{r} - \textbf{r}_i;
    \textbf{r}_{ij}, \left\{ \lambda \right\} ) \qquad \forall \quad r_{ij} < r_\text{SOAP},
\end{equation}
where $g$ is a smearing function (e.g., a Gaussian) centered at the position
of each atomic neighbor~\cite{bartok_2013,caro_2019}. We have collectively represented the
different SOAP hyperparameters, which determine the fine detail of the atomic representation,
by $\{ \lambda \}$.
The integral over relative orientations in \eq{SOAP_kernel} ensures the rotational invariance of this
kernel, which also satisfies the permutational and translational invariance criteria.
The integrand is raised to power $n\geq 2$ to retain all the angular information of
the original environments~\cite{bartok_2013}.
To make the integral computationally tractable, the
atomic density field is expanded in terms of a radial basis $\{g_n\}$ and
spherical harmonics $Y_{lm}$:
\begin{equation}
    \rho \ofr \equiv \sum_{nlm} c_{nlm} \, g_n(r) Y_{lm} (\theta,\phi),
\end{equation}
with the normalized power spectrum of the expansion coefficients $c_{nlm}$ giving the
SOAP vectors:
\begin{equation}
    p_{nn'l} = \sum_{m} c_{nlm} (c_{n'lm})^*, \qquad
    \textbf{q} = \frac{\textbf{p}}{\sqrt{\textbf{p} \cdot \textbf{p}}}.
\end{equation}
The final (normalized) form of the SOAP kernel between atomic
environments $i$ and $j$ is given by a dot product~\cite{bartok_2013,caro_2019}:
\begin{equation}
    k^{\text{SOAP}}(i,j) \equiv \mathbf{q}_i \cdot \mathbf{q}_j,
    \label{SOAP}
\end{equation}
where $i$ refers to the atom with density $\rho$ and $j$ to the atom with density $\rho'$ in
\eq{SOAP_kernel}. In this work we use the formulation of SOAP introduced
in Ref.~\cite{caro_2019}, which offers several computational advantages. SOAP is
particularly well suited for the representation of atom-based properties that depend on
the entire (many-body) local atomic environment.

\subsection{ML model of Hirshfeld volumes and gradients}

In our vdW model, SOAP descriptors are used to predict the effective Hirshfeld
volumes of the different atoms based on their local atomic environments. These volumes
can then be used to calculate the TS dispersion correction using \eq{TS}. The effective
Hirshfeld volume for atom $i$ is thus predicted by
\begin{equation}
    \nu_i = \sum\limits_{s \in S} \alpha_s |k^{\text{SOAP}}(i,s)|^{\zeta},
    \label{volumes}
\end{equation}
where $\alpha_s$ are the fitting coefficients obtained from the ML model and $S$
denotes the chosen sparse set of atomic environments, which is a subset of the whole
training set $T$. We implicitly assume $\delta = 1$ [cf., \eq{kernel_eq}].
The parameter $\zeta$ is empirical and takes small positive values $\zeta \geq 1$.
It is used to make the kernels sharper to effectively emphasize the differences between
environments and to accentuate the sensitivity of the kernel to changing atomic
positions~\cite{bartok_2013}. To further improve the fitting, the reference DFT data
can be shifted by approximately the mean of the whole database:
\begin{equation}
    \nu_i = \nu_i^{\text{DFT}} - \nu_0, \quad \nu_0 = \frac{1}{N}\sum\limits_{i=1}^N \nu_i^{\text{DFT}}.
    \label{v_shift}
\end{equation}
Subtraction of $\nu_0$ is done for the training stage (so that the quantity to be
learned is smoother) and then added back at the prediction stage.
With these predicted Hirshfeld volumes we can calculate the pairwise dispersion
energies using the TS dispersion correction given by \eq{TS} and \eq{smooth}.

The dispersion forces are given as the negative gradients of the dispersion energy:
\begin{equation}
    f_k^{\alpha} = -\frac{\partial E_{\text{TS}}}{\partial r_k^{\alpha}},
\end{equation}
where $k$ denotes the atom and $\alpha$ denotes the Cartesian coordinate. Here we give
the result of the differentiation (see SM~\cite{sm} for derivation details):
\begin{multline}
    f_k^{\alpha} = \sum\limits_{i}\frac{\partial \nu_i}{\partial
    r_k^{\alpha}}\sum\limits_{j\neq i} \Bigg[\frac{C_{6,ij}}{\nu_i}
    \frac{f_{\text{damp}}(r_{ij})}{r_{ij}^6} f_{\text{cut}}(r_{ij})
    \\
    -\frac{r_{ij}}{(r_{ij}^{\text{vdW}})^2}\frac{d}{s_\text{R}}f_{\text{damp}}(r_{ij})^2
    f_{\text{cut}}(r_{ij})
    \\
    \times \exp\left(-d\left(\frac{r_{ij}}{s_\text{R} r_{ij}^{\text{vdW}}}-1\right)\right)
    \frac{C_{6,ij}}{r_{ij}^6}\frac{1}{3 \nu_i^{2/3}} r_{i,\text{free}}^{\text{vdW}} \Bigg]
    \\
    + \sum\limits_{i}\delta_{ik}\sum\limits_{j\neq i}
    \frac{C_{6,ij}}{r_{ij}^7}f_{\text{damp}}(r_{ij})f_{\text{cut}}(r_{ij})(r_j^{\alpha}-r_i^{\alpha})
    \\
    \times \Bigg[\frac{6}{r_{ij}} - \frac{d}{s_\text{R} r_{ij}^{\text{vdW}}}f_{\text{damp}}(r_{ij}) 
    \exp\left(-d\left(\frac{r_{ij}}{s_\text{R} r_{ij}^{\text{vdW}}}-1\right) \right) \Bigg]
    \\
    + \sum\limits_{i}\delta_{ik}\sum\limits_{j\neq i} 
    \frac{C_{6,ij}}{r_{ij}^7}f_{\text{damp}}(r_{ij})(r_j^{\alpha}-r_i^{\alpha}) D_{ij}.
    \label{force_terms}
\end{multline}
where $\delta_{ik}$ is the Kronecker delta and $D_{ij}$ is defined below.
We note that the expression above includes terms whose differentiation is trivial,
but also terms which depend on the gradient of the Hirshfeld volume, which is a
functional of the electron density. Since this functional is non-variational with
respect to changes in the charge density (i.e., the $\nu_i$ are not obtained via
minimization, but simply by numerical integration), the Hellmann-Feynman theorem does
not apply. Consequently, most DFT codes do not report the contribution of the
gradients of the Hirshfeld volumes to the dispersion forces.

By contrast, in our method the gradient of the Hirshfeld volume with respect to
the position of atom $k$ is straightforward to compute in terms of kernel and
descriptor derivatives:
\begin{equation}
    \frac{\partial \nu_i}{\partial r_k^{\alpha}} = \sum\limits_{s\in S}
    \alpha_s \zeta |k(i,s)|^{\zeta-1}\mathbf{q}_s \cdot \frac{\partial\mathbf{q}_i}{\partial r_k^{\alpha}}.
\end{equation}
The derivatives of the SOAP descriptors are readily available from any code
used for regular cohesive energy GAP computations which is able to compute forces.
In our case, these derivatives are implemented in the GAP~\cite{bartok_2015}
and TurboGAP~\cite{caro_2019}
codes. Finally, the coefficient $D_{ij}$ that appears due to the derivative of
the smoothing function $f_\text{cut}(r_{ij})$ on the final line of Eq.~\eqref{force_terms}
is given by:
\begin{equation}
    D_{ij} =
    \begin{cases}
    -\frac{6}{d_{\text{b,o}}} (-r_{\text{b,o}}+r_{\text{b,o}}^2), &\text{if }
    r_{\text{c,o}} - d_{\text{b,o}} < r_{ij} \leq r_{\text{c,o}}  \\
    -\frac{6}{d_{\text{b,i}}} (r_{\text{b,i}}-r_{\text{b,i}}^2), &\text{if }
    r_{\text{c,i}} < r_{ij} \leq r_{\text{c,i}} + d_{\text{b,i}}  \\
    0, & \text{otherwise},
    \end{cases}
\end{equation}
with $r_\text{b,o}$ and $r_\text{b,i}$ given in Eq.~\eqref{smooth}.

\subsection{TurboGAP and QUIP/GAP implementations}

We have implemented the present methodology in the GAP/QUIP~\cite{ref_quip} and TurboGAP~\cite{ref_turbogap}
codes. To predict the dispersion energies and forces, the implementation uses
the fitting coefficients $\boldsymbol{\alpha}$ and the matrix of SOAP descriptors for the
sparse set $\mathbf{Q}_S$. Those are both precomputed at the training stage from
the DFT reference data of Hirshfeld volumes for different atomic environments and saved to a file,
which is read and stored in memory at the beginning of a new simulation. Given a new
atomic structure (a molecule or a solid), the Hirshfeld volumes are predicted for each atom
in the structure using their SOAP many-body descriptors, which are computed on the fly, and
the precomputed
$\boldsymbol{\alpha}$ and $\mathbf{Q}_S$. The predicted Hirshfeld volumes are given by
\eq{volumes}, which is used in matrix form in TurboGAP:
\begin{equation}
    \nu_* = (\mathbf{q}_* \mathbf{Q}_S^\intercal)^{\odot\zeta}\boldsymbol{\alpha} + \nu_0.
\end{equation}
Here we denote the effective Hirshfeld volume and the SOAP descriptor of a new atomic
environment with the subscript ``$*$''. The Hadamard operator $\odot$ indicates
element-wise exponentiation, in this case. If the forces are to be calculated, the
derivatives of the Hirshfeld volumes are calculated in the same part of the code, reusing
the different array quantities:
\begin{equation}
    \frac{\partial\nu_{*}}{\partial r_k^{\alpha}} = \left(\zeta(\mathbf{q}_{*}
    \mathbf{Q}_S^\intercal)^{\odot(\zeta-1)}\odot\boldsymbol{\alpha}\right) \mathbf{Q}_S
    \frac{\partial \mathbf{q}_*^\intercal}{\partial r_k^{\alpha}},
\end{equation}
where $\mathbf{q}_* \in \mathbb{R}^{1\times N_{\text{SOAP}}}$,
$\boldsymbol{\alpha} \in \mathbb{R}^{1\times N_S}$ and $\mathbf{Q}_S \in
\mathbb{R}^{N_S \times N_{\text{SOAP}}}$. Here, the second
Hadamard operator indicates element-wise multiplication.

Since the SOAP many-body descriptors only see the local environment within the SOAP cutoff
sphere of radius $r_{\text{SOAP}}$, the descriptor of atom $i$ also only has non-zero
SOAP gradients for atoms within that cutoff. The SOAP descriptor of atom $i$ gives the
gradients for a SOAP neighbor (within the SOAP cutoff) atom $k$ as
${\partial \mathbf{q}_i}/{\partial r_k^{\alpha}}$.
The cutoff for pairwise vdW interactions $r_{\text{cut}}$ is generally different and usually
larger than $r_{\text{SOAP}}$, because the vdW interaction is long ranged and SOAP descriptors
encode the local atomic environment only. It is technically possible, however, to define
them such that $r_{\text{SOAP}}$ is larger. Whichever cutoff radius is the larger one will
determine the neighbor list that needs to be built. If atom $i$ in the structure has $n_i$ neighbors
within $r_{\text{cut}}$ (counting also the atom itself), and assuming it is the largest cutoff,
the neighbor list will
have $N_{\text{pairs}} = \sum\limits_{i} n_i$ elements. The following quantities are then
calculated for all pairs of atoms $i$ and $j$ within the vdW cutoff radius:
\begin{align}
    C_{6,ij}, \quad f_{\text{damp}}(r_{ij}; r_{ij}^{\text{vdW}}, s_\text{R}, d),
    \quad r_{ij}^{-6}, \quad r_{ij}^{\text{vdW}},
    \\
    \exp\left(-d\left(\frac{r_{ij}}{s_\text{R} r_{ij}^{\text{vdW}}}-1\right)\right),
    \quad r_j^{\alpha} - r_i^{\alpha}
\end{align}
and stored in vectors of size $N_{\text{pairs}}$. By looping over the atoms $i$ and their
neighbors $j$, these quantities can be used to calculate the inner sums in \eq{force_terms},
such that we get an equation that reads
\begin{equation}
    f_k^{\alpha} = \sum\limits_{i} \frac{\partial \nu_i}{\partial r_k^{\alpha}} A_i
    + \sum\limits_{i}\delta_{ik} B_i.
\end{equation}
The explicit expressions for $A_i$ and $B_i$ can be retrieved by collecting the corresponding
terms in \eq{force_terms}.
Then, by looping over atoms $i$ and their SOAP neighbors $k$ again, we get the final dispersion
forces for the atoms. Here, the gradients of the effective Hirshfeld volumes are zero if $r_{ik} > r_{\text{SOAP}}$, and those terms can be skipped in the sum.
The first term involves the effect of the changing electronic density within the local
environment on the atoms in that environment. The second term is the long-range part that
also appears when neglecting the gradients of the Hirshfeld volumes. As we have discussed
earlier, this second term gives the same TS forces that are usually reported by DFT codes,
which often lack the interactions given by the first term altogether. We provide some illustrative
examples of this issue and how the present implementation overcomes it in Sec.~\ref{10}.

A reference implementation is provided in the QUIP package~\cite{ref_quip}, which together
with the GAP plugin also facilitates fitting the model of Hirshfeld volumes from a database
of atomic configurations. For a database containing $N$ atoms, with effective Hirshfeld
volumes obtained from DFT calculations (shifted by the approximate mean as in Eq.~\eqref{v_shift}) $\boldsymbol{\nu}$, the set $\{\mathbf{q}_i\}_{i=1}^N$
is first computed, of which $N_S$ representative points are selected as the sparse set.
Elements of covariance matrices $\mathbf{K}_{TS}$ and $\mathbf{K}_{SS}$
are computed using the kernel definition in \eq{kernel_eq}, where the subscripts
$T$ and $S$ indicate the full training set and sparse set of atomic environments, respectively. In matrix form these are given by:
\begin{equation}
    \mathbf{K}_{SS} = \left(\mathbf{Q}_S \mathbf{Q}_S^\intercal\right)^{\odot \zeta}, \quad \mathbf{K}_{TS} = \left(\mathbf{Q}_T \mathbf{Q}_S^\intercal\right)^{\odot \zeta}, \quad \mathbf{K}_{ST} = \mathbf{K}_{TS}^\intercal,
\end{equation}
where $\mathbf{Q}_T \in \mathbb{R}^{N_T \times N_\text{SOAP}}$ and $\mathbf{Q}_S \in \mathbb{R}^{N_S \times N_\text{SOAP}}$ have the SOAP descriptors as their rows.
The coefficients are obtained using the sparse Gaussian process regression~\cite{snelson06} as
\begin{equation}
    \boldsymbol{\alpha} = ( \mathbf{K}_{SS} + \mathbf{K}_{ST} \boldsymbol{\Sigma}^{-1}
    \mathbf{K}_{TS} )^{-1} \mathbf{K}_{ST} \boldsymbol{\Sigma}^{-1} \boldsymbol{\nu},
\end{equation}
where $\boldsymbol{\Sigma}$ is a diagonal matrix containing the regularization parameters
associated with each effective Hirshfeld volume observation.
To predict the vdW interaction terms in QUIP, we first obtain the atomic force constants
$C_{6,ij}$ and the vdW radii $r^{\mathrm{vdW}}_{ij}$, as well as their derivatives, if required,
with respect to the Cartesian coordinates of neighboring atoms.
For each atomic pair within the vdW cutoff, these terms are substituted in the TS expression
of the dispersion energy, while force and stress contributions are also accumulated.

The QUIP implementation, while less efficient than the alternative TurboGAP, offers
the possibility of fitting and using hierarchical ($\Delta$-learning) models, as well as
Python bindings via the quippy extension of QUIP and ASE~\cite{larsen_2017} integration.

For computational efficiency, the TurboGAP code offers the capability of reusing the
same SOAP descriptors that are used for constructing a cohesive energy GAP for the Hirshfeld
volume prediction. In practical terms, this means that the TS corrections as implemented
here can be applied on top a cohesive energy force field with only a small increment in
computational cost, of the order of 20-50\% extra CPU time
for the systems studied here with the C60 GAP. We note, however, that the
computational overhead depends on many factors, such as the effective range of vdW
interactions for the system under study, the ratio of vdW cutoff to SOAP cutoff (since
the scaling laws are different), the typical number of neighbors within the cutoff
sphere, or the relative cost of the kernel regression step compared to building
the descriptors.

\section{Simulations of carbon-based systems}

As a proof of principle for the new ML implementation of vdW interactions, we study the
phase diagram of the \csixty{} molecular system. The interaction between individual
\csixty{} molecules at low pressures is almost exclusively driven by vdW interactions,
whereas, as temperature and pressure increase, the role of repulsion and covalent
interactions (in the form of broken and created chemical bonds) becomes more prominent.
At very high pressure there is a transition from \csixty{} to amorphous carbon, whereas
at high temperature (but lower pressure) the \csixty{} molecules decompose to graphite.
Both transitions require a reactive force field to be characterized.
Thus, this is a problem that requires accurate description of both vdW interactions and
chemical reactions via a vdW-inclusive reactive force field. This combination of features
can currently only be delivered by \textit{ab initio} methods or the present generation
of ML potentials. At the time and length scales necessary to carry out a comprehensive
characterization of the phase diagram of \csixty, \textit{ab initio} methods are
prohibitively expensive, leaving ML force fields as the only viable option.

\subsection{C60 GAP force field}

\begin{figure}[t]
    \centering
    \includegraphics[width=\columnwidth]{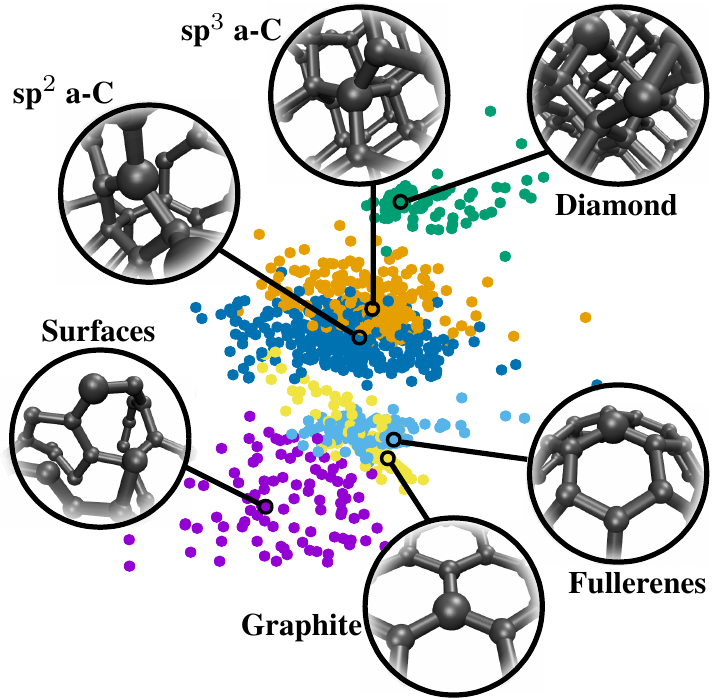}
    \caption{Overview of the atomic configurations present in the database, using a
    sparse set of 1000 structures. Their composition is captured in this 2D embedding,
    where the distances between points mimic the SOAP dissimilarities between the
    corresponding atoms. We used a soap\_turbo kernel with 4~\AA{} cutoff for the
    dissimilarity measure and a simple hierarchy of six clusters. Several representative
    structures are shown for reference.}
    \label{07}
\end{figure}

\begin{figure}[t]
    \centering
    \includegraphics[width=\columnwidth]{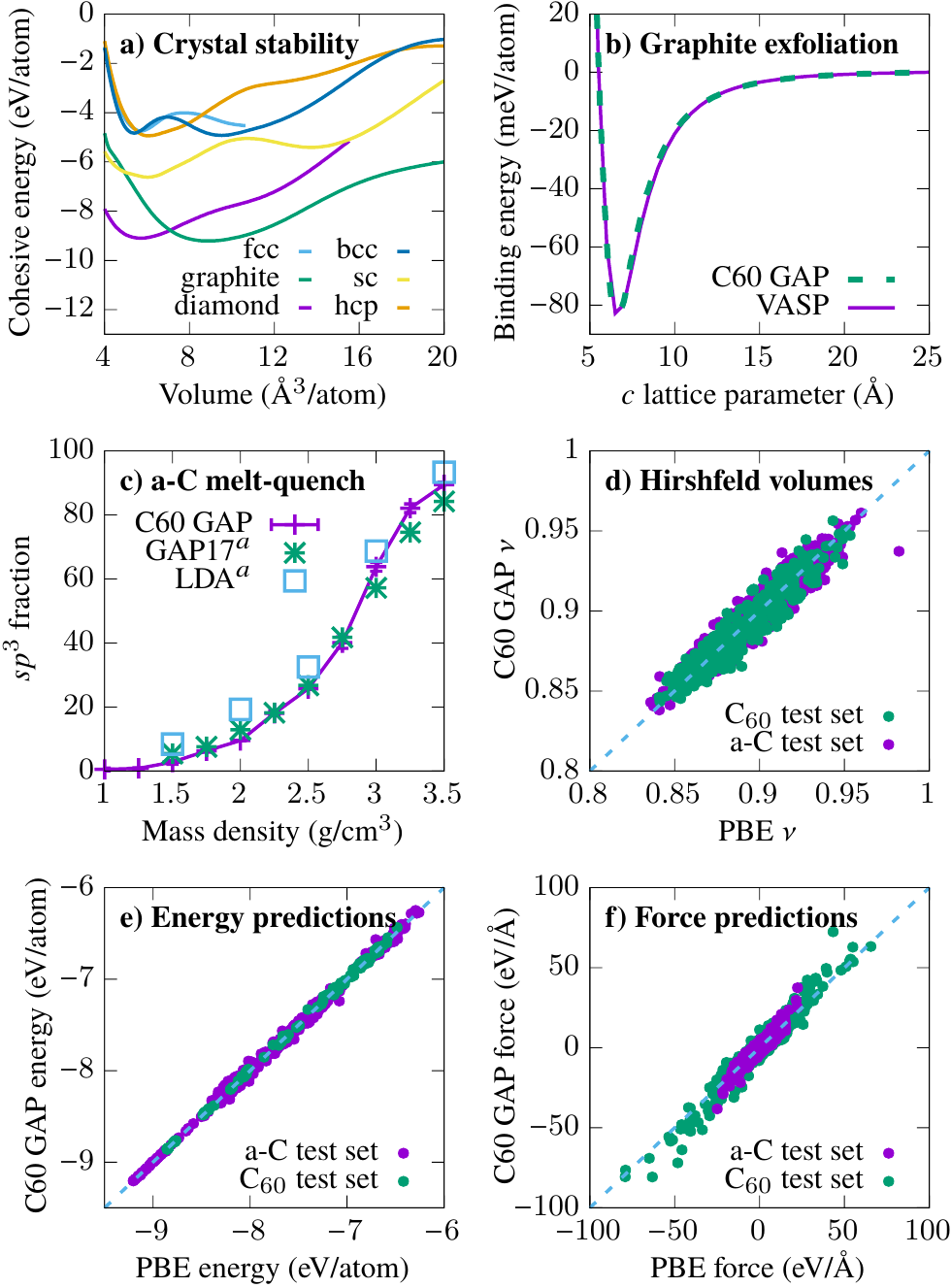}
    \caption{Basic tests of the C60 GAP. a) Cohesive energy predictions for
    six crystal structures vs. volume, showing the
    expected stability of diamond at high pressure and graphite at low
    pressure. All other phases are much higher in energy. b) Graphite
    exfoliation curve, compared to the VASP reference data, including TS
    corrections with a 50~\AA{} cutoff (the standard cutoff
    used by VASP). The graphite
    exfoliation curve shows sub-meV/atom agreement between the DFT reference
    data and the C60 GAP. c) a-C generation throughout different mass
    densities (ten 216-atom simulations per density value were used),
    closely following the strategy presented in Ref.~\cite{deringer_2017} and
    comparing to data reported therein. d) Predicted Hirshfeld volumes for
    a-C and \csixty{} test sets not included in the training set. The RMSEs
    are 0.0051 and 0.0046 for the a-C and \csixty{} test sets, respectively.
    e) Energies predicted by the C60 GAP (without vdW corrections) on the a-C
    and \csixty{} test sets vs the PBE predictions. We do not compare
    against the PBE+TS energies since the agreement between TS and our present
    methodology is fully captured by the Hirshfeld volume comparison in d);
    further comparison is
    provided in Sec.~\ref{10}. The RMSEs are 37 and 21 eV/atom for a-C and \csixty{}
    test sets, respectively. f) The same comparison as in e) but for forces
    in this case, with RMSEs of 1.17 and 1.69 eV/~\AA{} for a-C and \csixty{},
    respectively. $^a$Data from Ref.~\cite{deringer_2017}.}
    \label{01}
\end{figure}

To accurately describe the strong interatomic interactions in carbon materials, i.e.,
repulsion and covalent interactions, we constructed a new GAP force field for C, using the
structural database of the 2017 amorphous carbon (a-C) GAP of Deringer and
Cs\'anyi (GAP17)~\cite{deringer_2017} as a baseline. GAP17 contains different amorphous
and liquid C structures, dimer configurations, diamond, graphite and reconstructed
surfaces. It is thus a very solid general-purpose carbon potential that gives reasonable
results for most test carbon structures. However, it lacks fine resolution for some
applications. For the present study, we enhanced the structural database of GAP17 with
graphitic structures (exfoliation curves of graphite and bilayer graphene, and glassy
carbon), some reconstructed crystalline (diamond) surfaces, and plenty of \csixty{}
based structures, such as distorted \csixty{} monomers, undistorted \csixty{} dimers
and colliding \csixty{} dimers. The latter are useful to correctly describe the
decomposition of \csixty{} molecules at high pressure. We also included some C trimer
structures to improve the stability of the 3b GAP terms under high compression. For the
Hirshfeld volume fit, the GAP17 database and the new \csixty{} structures were used.
The GAP17 and new database were computed at the DFT level of theory with the
PBE~\cite{perdew_1996} functional with VASP~\cite{kresse_1996,kresse_1999}. The reference
TS calculations (including the Hirshfeld volume computations) follow the implementation
in VASP by Bu{\v{c}}ko \textit{et al}.~\cite{bucko_2013}. The composition of the database
of atomic structures can be visualized in \fig{07}, where we use a new method to embed
high-dimensional data in two dimensions, based on a hierarchical combination of
cluster-based data classification and multidimensional
scaling~\cite{hernandez-leon_2021,ref_clmds}. This is a popular tool for visualizing
structural databases in the context of ML applied to the study of atomic
systems~\cite{de_2016,caro_2018c,deringer_2020,cheng_2020}.

\begin{table}[b]
\caption{Timings for the C60 GAP, computed on a test system made up of 64 \csixty{}
units at 10~kbar; see Sec.~\ref{06} for a discussion on the structure. The timings
exclude the time spent on neighbors lists, reading input files, etc., they only
report the evaluation of energy and forces. The timings are obtained running
TurboGAP on a compute node equipped with two Intel Xeon Gold 6230
CPUs at 2.1~GHz clock speed (20 physical cores per CPU). We note that
these are CPU times (real time times number of cores), that is the reason why the timings
increase with more cores. An example simulation of 256 \csixty{} molecules (15360 atoms)
over 100~ps
of MD with 1~fs time step, run on 40 CPU cores with the vdW cutoff radius set to 20~\AA{},
would take 1280 CPUh, or 32h of real time.}
\begin{ruledtabular}
\begin{tabular}{r | c c c c}
\# cores & \multicolumn{4}{c}{CPU time (ms/atom/MD step)} \\
& No vdW & GAP+Hirshfeld & vdW (10~\AA{}) & vdW (20~\AA{}) \\
\hline
1 & 1.29 & 1.49 & 1.58 & 2.25 \\
5 & 1.48 & 1.57 & 1.64 & 2.23 \\
10 & 1.62 & 1.70 & 1.79 & 2.42 \\
20 & 2.11 & 2.18 & 2.31 & 3.07 \\
40 & 2.13 & 2.24 & 2.65 & 3.00 \\
\end{tabular}
\end{ruledtabular}
\label{05}
\end{table}

The new C60 GAP potential uses 2b descriptors with 4.5~\AA{} cutoff, 3b
descriptors with 2.8~\AA{} cutoff and soap\_turbo mb descriptors, as available in
the GAP code, with 4.5~\AA{} cutoff. The potential is freely available from the
Zenodo repository~\cite{muhli_2021} and can be used to run simulations with QUIP/GAP
and TurboGAP. Besides the new vdW capabilities, this potential also incorporates
a ``core potential'' term that ensures accurate simulation also at very close
interatomic separations. E.g., the potential should remain stable for simulations
of atomic collisions up to 1-2 keV, although quantitative accuracy at these
energies has only been specifically enforced for the C dimer. Therefore, even
though we introduce the potential as tailored for \csixty{} simulation, it should
also be regarded as a high-quality general-purpose potential to model carbon
materials, graphitic carbons in particular. Basic tests of this potential are
summarized in \fig{01}, showing very good performance across various applications.
We emphasize that the errors for the a-C tests are actually smaller than those
reported for GAP17~\cite{deringer_2017}, on which the C60 GAP is based. This
further highlights C60 GAP as an excellent general-purpose carbon potential,
beyond being specifically designed to accurately model \csixty{} systems.

Table~\ref{05} shows timings for the present C60 GAP potential. As can be seen,
the potential runs rather fast for the tested system and addition of Hirshfeld
volume prediction has a very small overhead of circa 5\%. Adding vdW corrections
results in an additional CPU cost that depends strongly on the vdW cutoff. For
this test system, the overall overhead of adding vdW corrections, including
Hirshfeld volume prediction, is circa 20\% with a 10~\AA{} vdW
cutoff. This number
grows quickly to circa 50\% with a 20~\AA{} cutoff. These two values will delimit
the range of practically relevant cutoffs for most systems. In particular for carbon
materials, a vdW cutoff of $\sim 15$~\AA{} can be considered sufficient, since the
pairwise vdW interactions decay very rapidly beyond this point [see \fig{01} b) for
the case of graphite and the SM~\cite{sm} for a thorough numerical analysis
in \csixty{} and a-C]. We find that the TurboGAP code scales well with the number of
CPU cores, and that the addition of vdW corrections does not seem to affect the
scaling behavior. For the test shown in Table~\ref{05}, the rule of thumb is that
as the number of CPU cores increases by a factor of 2, the calculation runs about a
factor of x1.8 faster. In general, how far the calculation scales will depend on the
number of atoms being simulated. For small systems, scaling is limited to a couple of
compute nodes. For very large systems, more CPU cores can be used. We managed to get
good scaling for a 1M-atom graphitic carbon system up
to 2048 MPI processes (on as many physical CPU cores). Efforts on improving the
software and computational performance are ongoing, and will be reported elsewhere.

\subsection{Hirshfeld volumes: locality and learning rates}

\begin{figure}[t]
    \centering
    \includegraphics[width=\columnwidth]{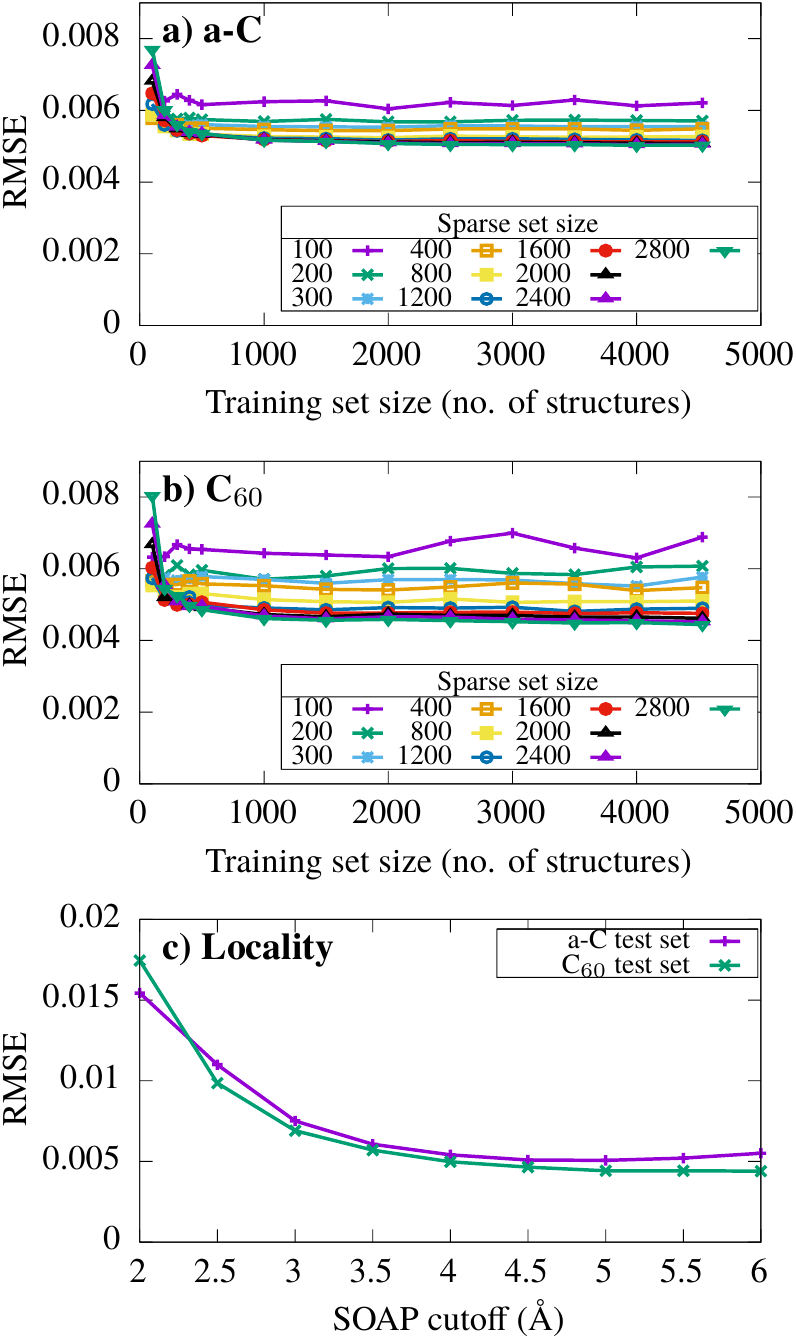}
    \caption{Hirshfeld volume (a-b) learning rates and c) locality tests for
    amorphous carbon (a-C) and $\text{C}_{60}$ test sets. The RMSEs are for
    predicted effective Hirshfeld volumes that are usually slightly smaller than
    one (the average of the training set is approximately $0.9$). See \fig{01}
    d) for a representative range of variation in these data sets.}
    \label{fig:RMSE}
\end{figure}

The model involves several hyperparameters that can improve the predictions
but also have a huge impact on the required computational time. Two of these
parameters are the SOAP descriptor cutoff radius and the size of the sparse
set. The former gives the locality of the physical parameters that the
model is predicting and the latter determines the learning rate of the model.
Fig.~\ref{fig:RMSE} shows the root-mean-square error (RMSE) of the
predicted effective Hirshfeld volumes for amorphous carbon and $\text{C}_{60}$ test sets as a function of
SOAP cutoff radius and the sparse set size. In \fig{fig:RMSE} a) and b), for a-C and
$\text{C}_{60}$ test sets respectively, it can be seen that the errors decrease
quickly as the size of the training set or the sparse set is increased, but
the model accuracy saturates such that increasing the sparse set size is not beneficial
after a certain point due to the increased computational cost. The locality test
in \fig{fig:RMSE} c) shows that for the a-C test set there is an increase in
error after a certain SOAP cutoff. The descriptors of two different amorphous
carbon structures can be very different from one another if a large cutoff is
used while the atoms far away from the central atom do not affect the physics of
the system significantly. On the other hand, for the $\text{C}_{60}$ molecules
the descriptors can be quite similar even for large cutoff radii. This is also
the reason why the error has not yet reached the minimum for the $\text{C}_{60}$
test set and a larger cutoff could be used.

\subsection{Missing contribution to the forces in DFT codes}\label{10}

The effective Hirshfeld volumes are non-variational functionals of the charge
density and
calculating their gradients thus requires the calculation of the charge
density gradients. The DFT code we use as our reference method lacks the
implementation of these charge density gradients and the dispersion forces only
contain the terms of Eq.~\eqref{force_terms} that do not include the derivatives of
the Hirshfeld volumes. Within the GAP methodology, however, the calculation of
the gradients of the Hirshfeld volumes is straightforward and these terms can therefore
be included in the model.

\begin{figure}
    \centering
    \includegraphics[width=\columnwidth]{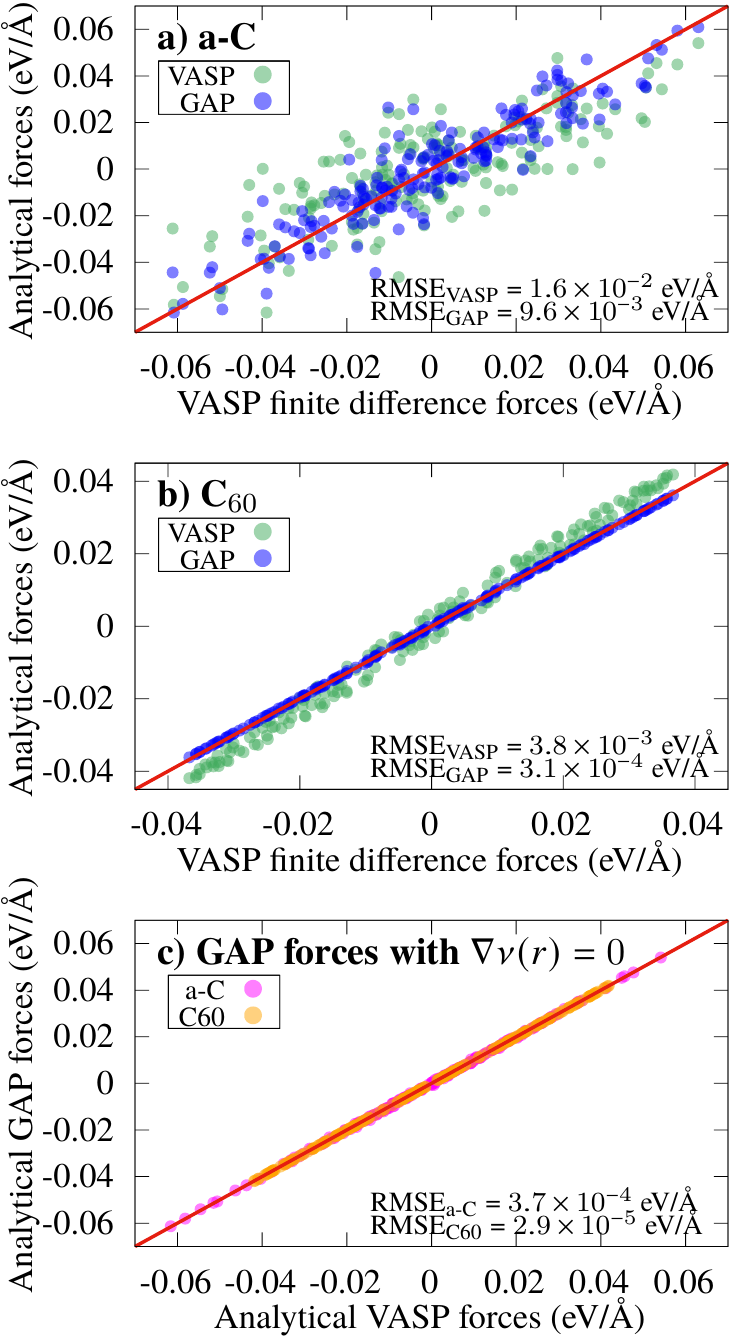}
    \caption{a) Comparison of dispersion forces ($x$, $y$ and $z$ components in
    the same panel) calculated analytically using the TurboGAP implementation and with
    the FD method from the dispersion energies calculated with VASP.
    Here the structure is amorphous carbon with 64 atoms in the unit cell and the
    cutoff radius for vdW interactions is 50~\AA. The VASP analytical forces show
    approximately 50~\% larger RMSE than the analytical GAP forces, when both are
    compared to the finite difference (which we treat as the reference here).
    b) Comparison of dispersion forces for a slightly distorted $\text{C}_{60}$
    structure with a vdW cutoff radius of 8~\AA{} to only include atoms in one
    $\text{C}_{60}$ molecule.
    The RMSE is an order of magnitude larger for VASP when compared to the GAP forces.
    c) Analytical GAP and VASP dispersion forces for amorphous carbon and
    $\text{C}_{60}$ structures. Here we have switched off the Hirshfeld volume
    gradient terms to show that we get excellent agreement with VASP dispersion
    forces, which do not include these gradients in any way.}
    \label{fig:forces}
\end{figure}

In Fig.~\ref{fig:forces} we demonstrate that our model is capable of reproducing
the dispersion forces calculated using the DFT method and that the missing contribution
can actually be of the same magnitude as the dispersion forces. Furthermore, we show
that the analytical dispersion forces that are produced by our model agree well with
the forces that we have calculated using the finite difference (FD) method from the DFT
total energies. For these calculations, we used an amorphous carbon structure with 64
carbon atoms in the unit cell and a distorted $\text{C}_{60}$ molecule. For the latter,
we only included the dispersion interactions within the molecule by making the
simulation box large enough and the vdW cutoff radius small enough.
For the former, we chose a vdW cutoff radius of 50~\AA{}, which is the
default cutoff used by VASP. We have included an analysis of the effect of vdW radius
on the dispersion forces in the SM~\cite{sm}.

For amorphous carbon, the RMSE is approximately 50\% larger for the VASP analytical
forces than it is for the analytical forces calculated with GAP, when compared to the
FD forces that we have calculated with VASP. For the distorted
$\text{C}_{60}$ structure the errors are an order of magnitude larger for VASP. The
database contains a number of different $\text{C}_{60}$ structures which enables
accurate interpolation with GAP. For the \csixty{} molecule the contribution from
the Hirshfeld volume gradients to the vdW forces is smaller, compared to a-C, thus
making the absolute and relative errors smaller too.

In Fig.~\ref{fig:forces}~c) we show that when we set the Hirshfeld volume gradients
to zero, effectively removing the first term of Eq.~(\ref{force_terms}), we obtain
almost perfect agreement with the analytical dispersion forces given by VASP. GAP is
thus able to closely reproduce VASP dispersion forces and, for structures
that are well represented in the database, it can indeed predict forces that are
\textit{more accurate}
than those produced by the reference method. This is one of the strongest results
in this paper, and we expect it to generalize to other problems where computation of
gradients is hindered by complications in the implementation.

\subsection{Structural transitions in \csixty{}}\label{06}

\csixty{} has been extensively studied in the
literature~\cite{quo_1991,johnson_1992,axe_1994,sundqvist_1999}, see, e.g.,
Refs.~\cite{pei_2019} and \cite{sundqvist_2021} for recent overviews. There is
fundamental interest, since \csixty{} is the molecule with the highest symmetry: 120
symmetry operations altogether (each an element in the icosahedral symmetry group),
including 60 rotational symmetry operations. But there is also a strong interest
in using \csixty{} precursors to synthesize novel forms of carbon by applying
different heat and pressure treatments~\cite{pei_2019,sundqvist_1999}.

Individual \csixty{} molecules are expected to behave similarly to soft spheres that
interact among each other via vdW forces at long distances. When the \csixty{}
molecules get close enough that the constituent carbon atoms ``see'' individual C-C
interactions (2b, 3b and mb), rather than an effective spherically symmetric potential
centered on the center of mass (CM) of the \csixty{} molecule, the situation becomes more
complicated. In particular, we are interested in finding out deviations from ideal
Lennard-Jones (LJ) fluid behavior at low pressure and in working out the thermodynamic
conditions for the transition from a \csixty{} fluid to amorphous carbon (a-C) and
liquid carbon (l-C) at high pressure and high temperature, respectively.
All the simulations in this section use the new C60 GAP and TurboGAP, a cutoff for vdW
interactions of 20~\AA{} and $s_\text{R} = 0.893$, which was optimized for C as detailed
in the SM~\cite{sm,tkatchenko_2009,jurecka_2007,marom_2011}.

\subsubsection{Clustering at low pressure}

\begin{figure*}[t]
    \centering
    \includegraphics[width=\textwidth]{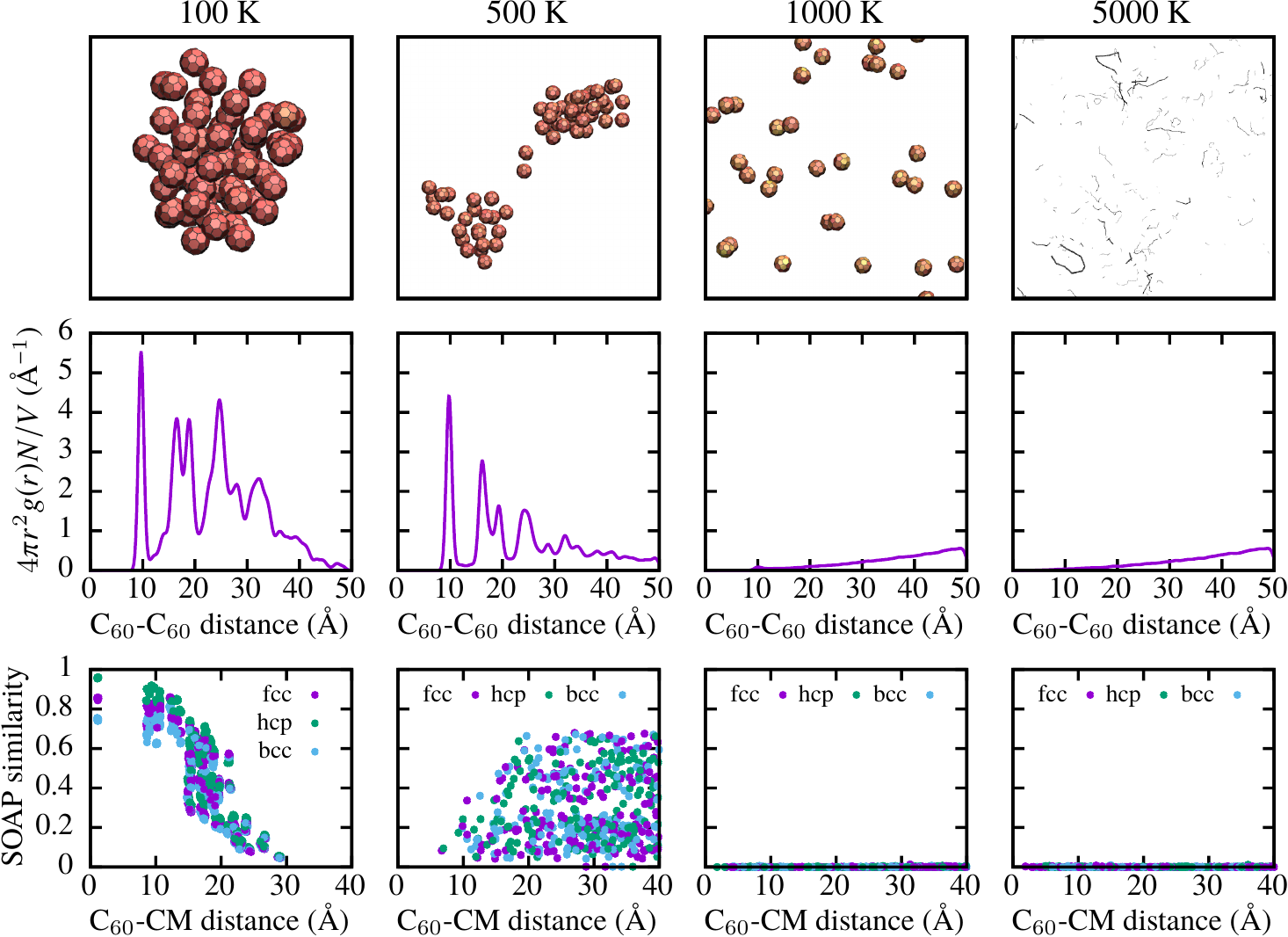}
    \caption{Top row: different structures resulting from the MD simulations; at low
    temperature, the \csixty{} units coalesce to form clusters, whereas as the
    temperature is raised to around and above 500~K these clusters become unstable
    and there is a transition
    to a \csixty{} gas first, and then to an atomic C gas. Middle row: radial
    distribution function (RDF) $g(r)$ for the center of mass (CM) representation of the
    \csixty{} molecules (i.e., each \csixty{} unit is mapped onto its CM and the CMs
    are used to compute the RDF). The RDF is shown, for convenience,
    multiplied by the factor required for
    the integral of the curve from 0 to $R$ to yield the number of \csixty{} units
    at $R$. Bottom row: SOAP similarity measures between the CM representation and
    reference simple crystal lattices, plotted as a function of distance between the
    CM of the \csixty{} units and the overall CM of the system. Here the kernel in
    \eq{SOAP} is raised to the power of 4 to improve visualization of the difference
    among crystal structures. Note that the SOAP descriptors used in this
    analysis are computed for the CM representation, not the full atomic
    representation.}
    \label{02}
\end{figure*}

At low pressure and low temperature, real systems that behave close to ideal LJ fluids,
such as Ar and other noble gases, tend to cluster into closed-packed
structures~\cite{rytkonen_1998,schwerdtfeger_2006}. To elucidate, first, the existence
and, subsequently, the structure of \csixty{} clusters at low pressure and temperature,
we carried out molecular dynamics (MD) simulations with our new C60 GAP potential and
the TurboGAP code. We simulated systems made up of 64 \csixty{} molecules (a total of
3840 C atoms) in large cubic simulation boxes of size $150 \times 150 \times 150$~\AA$^3$
under periodic boundary conditions (PBC). The positions of the initial configurations
were generated by randomizing the position of the CM of each \csixty{} molecule
while avoiding two \csixty{} units from coming too close to one another. We
scanned the temperature range from 10~K up to
5000~K, using a Berendsen thermostat~\cite{berendsen_1984} with time constant 100~fs
and time steps for the MD integration of 2~fs below 500~K and 1~fs at and above 500~K.
Each individual MD simulation at a given temperature is run for {500,000} time steps, i.e.,
1~ns below 500~K and 500~ps at and above 500~K. Structural analysis is performed after
the structural indicators, as described below, have stabilized. This means that we only
perform structural analysis on the final portion of the 1~ns or 500~ps trajectories.

To characterize the structure of the clusters we tried using the popular Steinhardt
parameter analysis~\cite{steinhardt_1983,yu_2014}, but this proved to be insufficient
to clearly resolve the structure of the \csixty{} clusters. Therefore, we resorted to
a more sophisticated analysis based on mb SOAP descriptors~\cite{bartok_2013,de_2016},
where the same kernel used in \eq{SOAP} to build GAP potentials is used to characterize
the similarity between atomic environments (see Ref.~\cite{caro_2020c} for a recent
example in carbon materials). For that purpose, we mapped each \csixty{}
unit into its CM and computed the radial distribution function (RDF) of the CMs from
our MD data. From the location of the first peak in the RDF, we inferred the typical
CM-CM first-neighbors distances in the \csixty{} fluid, and constructed
body-centered cubic (bcc), face-centered cubic (fcc) and hexagonal close-packed (hcp)
reference lattices. We then compared the SOAP descriptors of the \csixty{} CMs
in the simulated
clusters to the reference lattices, obtaining numerical scores for each of them. This
allows us to establish whether a given \csixty{} cluster resembles a bcc, fcc or hcp
structure the most. All the results for these low-pressure simulations are summarized
in \fig{02}.

If \fig{02}, the top row shows the structures resulting after equilibration for
selected temperatures: 100, 500, 1000 and 5000~K. We additionally performed
simulations at 10, 20, 50, 200 and 2000~K (not shown). However, the most representative
situations are those depicted in the figure. In particular, below 500~K we
observed that the individual \csixty{} units spontaneously coalesce to form
compact clusters. The characteristic distance between
\csixty{} units is about 9.7~\AA{}, and the structures are clearly ordered,
as evidenced from the RDF plot in the middle
row (note the RDF is computed for the \csixty{} CMs, not the individual C
atoms). The bottom row shows the similarity analysis between the actual
cluster and the reference fcc, hcp and bcc lattices. The similarity measures,
computed for each and all \csixty{} units in the system, are plotted as a function
of the distance between the CM of the entire system and the CM of the
corresponding \csixty{} unit. This is necessary since as one moves towards
the surface of the cluster the resemblance to an extended crystal is necessarily
lost. From this SOAP analysis, we conclude that the structure of the clusters
resembles a hcp lattice arrangement, and more so than fcc or bcc.
We should note, however, that at these low pressures and temperatures
the error in the TS vdW correction due to the missing many-body effects could
become significant. The importance of many-body effects on \csixty{} clustering
should be assessed when a computationally affordable implementation becomes
available. Our group is currently working on extending the present model to account
for these missing terms.

At around 500~K there is a transition between tightly and loosely connected
clusters, with the single cluster that forms at lower energies now splitting
into two subunits of similar size. The SOAP similarity analysis does not provide
anymore a clear preference for a specific crystal-like structure, although the RDF
still shows a clearly orderly arrangement.
Above 500~K the structure transforms into a gas and the vdW interactions are no
longer able to keep the \csixty{} molecules bound together. The individual \csixty{}
units remained chemically stable at 1000 and 2000~K (not shown), but they broke apart
into their atomic constituents at 5000~K. At this temperature the structure turns
into loosely connected open C chains. At this point the system does not retain any
memory of having originally been made of \csixty{} units.

\subsubsection{Solid phase at ambient conditions}

\begin{figure}[t]
    \centering
    \includegraphics[width=\columnwidth]{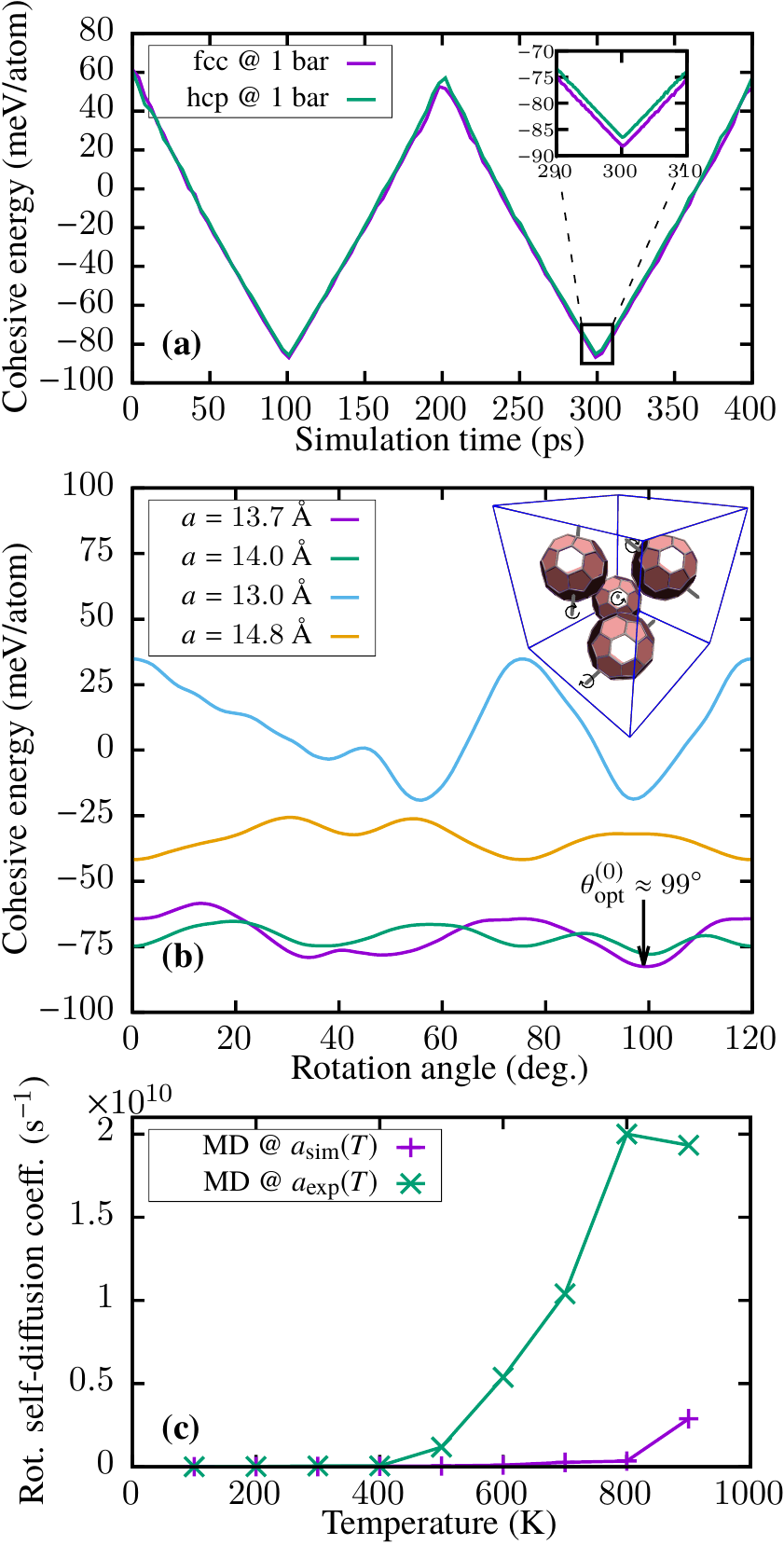}
    \caption{a) Cohesive energy (referenced to that of an isolated \csixty{}) for
    32-molecule systems in fcc and hcp structures as the temperature is cycled from
    1000~K to 10~K and back to 1000~K in 200~ps cycles, at $P = 1$~bar. b) Potential
    energy surface of a 4-molecule fcc unit cell as a function of the concerted
    rotation of angle $\theta$; see text for details. c) Rotational self-diffusion
    coefficient computed for the 32-molecule fcc system at different temperatures
    and 1~bar.}
    \label{08}
\end{figure}

At near ambient conditions, solid \csixty{} is experimentally known to adopt
an fcc structure~\cite{quo_1991,axe_1994}. There is also an experimental
observation of a transition from a rotationally
oriented phase, where the \csixty{} units are able to librate about equilibrium
positions and orientations but not freely rotate, to an orientationally disordered
structure where the \csixty{} can rotationally diffuse~\cite{axe_1994}.
The temperature of this
phase transition has been established experimentally at around 260~K. In this section
we try, with mixed success, to reproduce these experimental results, which are a
very stringent test
of the quality of our model due to the small magnitudes of the energies involved.

All the results of this section are summarized in \fig{08}. First, we try to elucidate
the crystal structure predicted by the C60 GAP for \csixty{} at ambient conditions.
\fig{08}~a) shows a series of successive heating and cooling cycles that we applied
to 32-molecule samples arranged in fcc and hcp lattices, coupled to a barostat at
1~bar and a variable thermostat that changes the temperature from 1000~K down to 10~K
and back up in 200~ps cycles. The figure shows that fcc is consistently lower in energy
than hcp at these conditions. For the low temperature structure (at 10~K), our results
indicate a cubic lattice constant for fcc of 13.7~\AA{}, and a cohesive energy of the order
of only 1.67~meV/atom (100~meV/molecule) lower (more stable) than hcp. Note that we fixed the
$c/a$ lattice parameter ratio of hcp to the ideal value during the simulation; it is
therefore possible that the energy difference might decrease even further if allowing for
non-ideal $c/a$ ratios (i.e., anisotropic barostating). These tiny energy differences, which
are one order of magnitude lower than typical available kinetic energies per degree of
freedom at room temperature, explain why under different conditions hcp may become more
stable than fcc. We found this to be the case for clusters in the previous section and
will also observe this behavior at high pressure and temperature in the next section.

Another intriguing aspect of the structure of \csixty{} at ambient conditions is the
orientationally ordered to orientationally disordered phase transition~\cite{quo_1991},
which has been thoroughly characterized in Ref.~\cite{axe_1994}. Namely, below the transition
temperature, experimentally observed at around 260~K, the four molecules at the fcc lattice
sites are rotationally ``locked'' in place. Above the transition temperature, they are free to
rotate about the fcc lattice sites. The specific orientation at low temperature can be
retrieved by a concerted rotation of all four molecules, by the same angle $\theta$, around
four different local axes (going though each molecule's center of mass). These axes are
indicated in the inset of \fig{08}~b), and in more detail in Refs.~\cite{axe_1994,sundqvist_2021}.
In that
figure we show the potential energy surface as a function of $\theta$. We find, for the
C60 GAP optimal lattice parameter, $a = 13.7$~\AA{}, that there is indeed a local minimum
at $\theta = 99^\circ$, very close to the experimental value of $98^\circ$. We also verified with
stochastic sampling (1 million random orientations of the four molecules around the fcc lattice
sites), that this is also probably the global minimum. We note, however, that this analysis
is based on a rigid molecule picture. Indeed, \fig{08}~a) shows that a lower energy
configuration can be achieved if the molecules are allowed to deform elastically (as is the
case in the course of MD). We also note the very strong dependence of the energy profiles
in \fig{08}~b) on the lattice parameter. At the low-temperature experimental value,
$a \approx 14$~\AA{}, this minimum becomes less stable compared to other local minima.
 
Finally, we tried to elucidate the rotational ordered-to-disordered phase transition by
calculating the rotational self-diffusion coefficient $D_\text{rot}$ for 32-molecule fcc
systems as a function of temperature, within 20~ps MD trajectories at each $T$.
Within the Stokes-Einstein
picture, $D_\text{rot}$ is expected to grow linearly in $T$ for freely rotating spherical
molecules. As expected for \csixty{} at low temperature, we find that this is not the
case in \fig{08}~c). There is clearly a transition
between a rotationally locked phase at low temperature to a rotationally diffuse phase at
high temperature, with exponential growth of $D_\text{rot}$, which increases by
several orders of magnitude, in between. This growth of $D_\text{rot}$ over a range of
temperatures is qualitatively in agreement with the existence of a range of temperatures
over which \csixty{} transitions from fully orientationally ordered to fully orientationally
disordered. Within that range, from $\sim 90$~K to 260~K, certain fractions of
molecules attain one of two possible
specific relative orientations (so-called ``P orientation'' and ``H orientation''),
the precise fraction depending on $T$ and $P$~\cite{sundqvist_2021}.
Again, we find in \fig{08}~c) that there is a strong dependence
of the specific characteristics on the lattice parameter. When we adjust the simulated lattice
parameter to the experimental one (which involves an increase of about 2.7\% in the
lattice parameter), the transition temperature goes down by about 300~K, from $\sim 800$~K
to $\sim 500$~K. Still, this temperature is far away, quantitatively, from the
experimental observation of the phase transition. This may be indicative,
together with the underestimation of the lattice parameter, that our approach may be
overestimating the vdW interaction. We remain, nonetheless, very satisfied with the overall
performance of the present approach, which qualitatively reproduces much of the fine detail of
the structure of \csixty{}, especially given the tiny energy differences involved.

\subsubsection{High pressure/high temperature phase transitions}

\begin{figure*}[t]
    \centering
    \includegraphics[width=\textwidth]{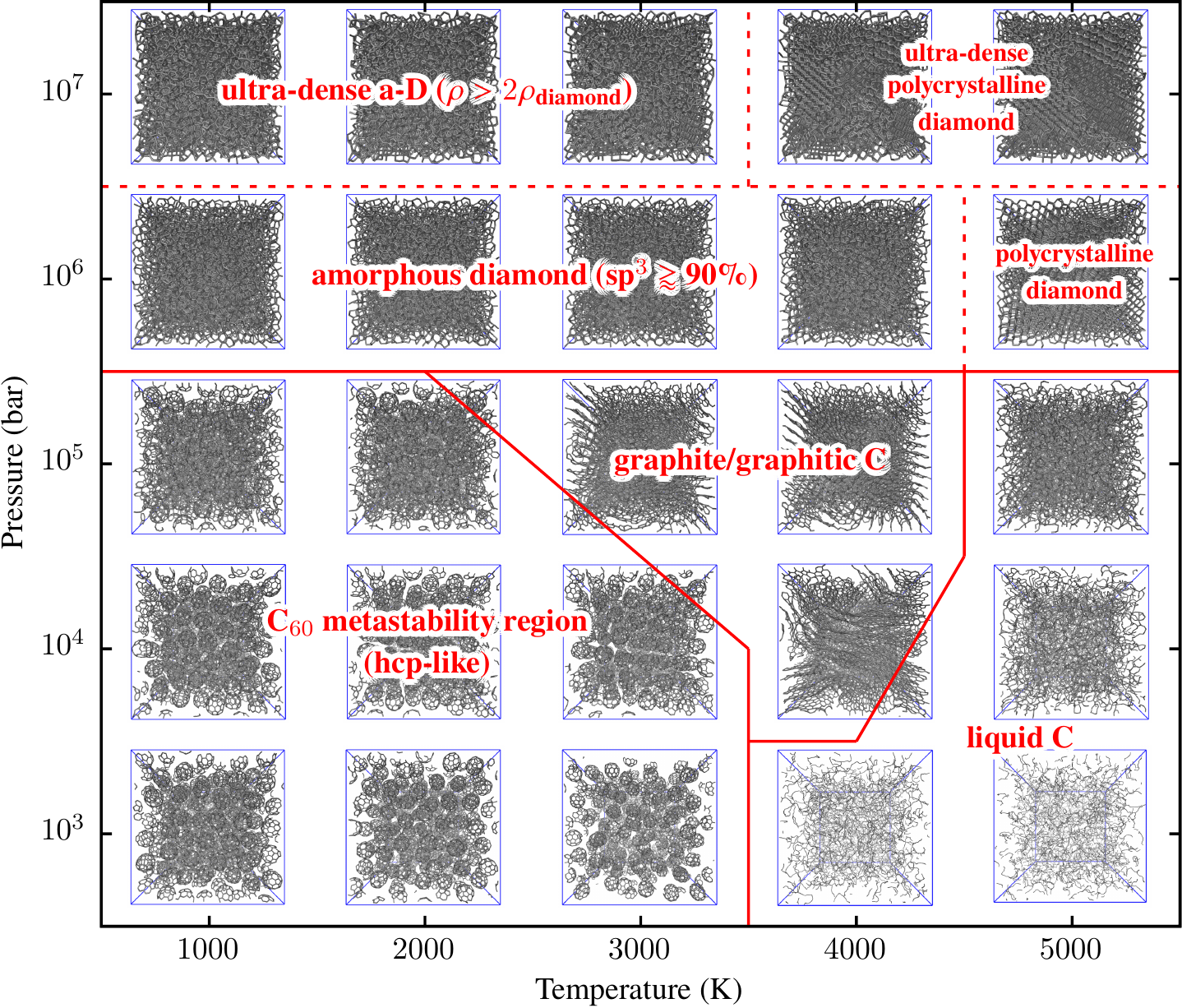}
    \caption{Metastable phase diagram for the \csixty{} system at high pressure (at and above
    1~kbar). Transitions to other allotropes of C from the \csixty{} seed can be observed
    as temperature, pressure, or both are increased. See text for a detailed discussion.
    The atomic structures were drawn with VMD~\cite{humphrey_1996,ref_vmd} and structure
    manipulation and analysis were carried
    out with ASE~\cite{larsen_2017} and different in-house codes.}
    \label{04}
\end{figure*}

\begin{figure}[t]
    \centering
    \includegraphics[width=\columnwidth]{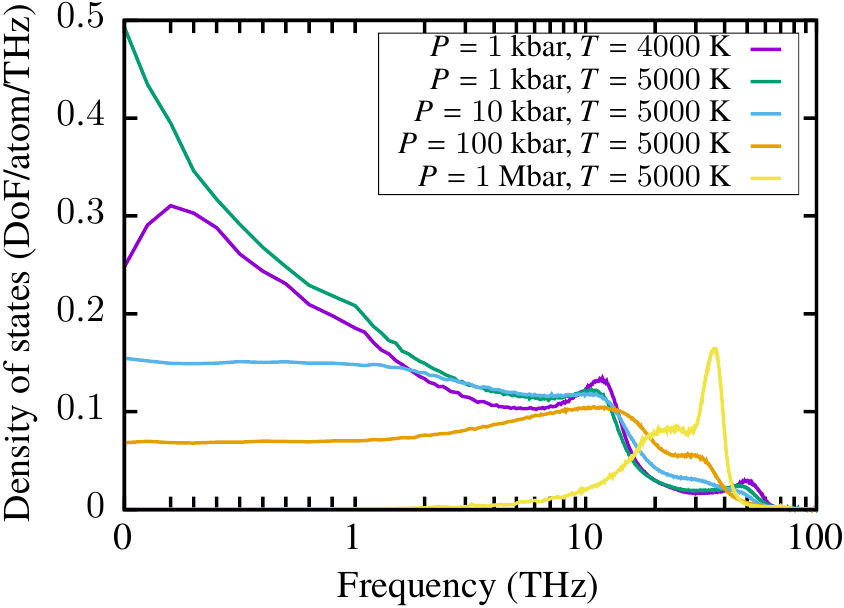}
    \caption{Vibrational density of states of all the liquid structures identified in
    \fig{04} plus the one identified as ``polycrystalline diamond'', which is used as
    a reference for a solid.}
    \label{09}
\end{figure}

\begin{table}[b]
\caption{Selection of structural indicators for the \csixty{} system at high pressure:
i) Approximate location of the first RDF peak in the CM representation of the
\csixty{} condensed phase (solid and liquid, i.e., excluding clusters at low pressure)
at different temperatures and pressures. ii) Mass densities for the different structures
depicted in \fig{04}.}
\begin{ruledtabular}
\begin{tabular}{r | c c c c c}
\multicolumn{6}{c}{RDF first peak} \\
& 1000~K & 2000~K & 3000~K & 4000~K & 5000~K \\
\hline
1~kbar & 9.85~\AA{} & 10.13~\AA{} & 10.45~\AA{} & N/A & N/A \\
10~kbar & 9.71~\AA{} & 10.03~\AA{} & 10.12~\AA{} & N/A & N/A \\
100~kbar & 8.86~\AA{} & 8.96~\AA{} & N/A & N/A & N/A \\
$\ge 1$~Mbar & N/A & N/A & N/A & N/A & N/A \\
\hline
\multicolumn{6}{c}{Mass density (g/cm$^3$)} \\
& 1000~K & 2000~K & 3000~K & 4000~K & 5000~K \\
\hline
1~kbar & 1.57 & 1.27 & 0.91 & 0.42 & 0.33 \\
10~kbar & 1.68 & 1.61 & 1.47 & 1.77 & 1.23 \\
100~kbar & 2.09 & 2.09 & 2.41 & 2.40 & 2.41 \\
1~Mbar & 3.89 & 3.90 & 3.90 & 3.90 & 3.96 \\
10~Mbar & 7.23 & 7.21 & 7.16 & 7.17 & 7.16 \\
\end{tabular}
\end{ruledtabular}
\label{03}
\end{table}

We have further studied the structure of the \csixty{} fluid under high pressure
conditions. In particular, we are interested in demarcating the phase boundary
between a molecular fluid/solid made up of individually stable \csixty{} units, on
the one hand, and a-C and l-C, on the other. In these simulations, we scanned the
temperature range where we expect to find the phase boundary, and ran MD simulations
at 1000, 2000, 3000, 4000 and 5000~K. We scanned the range of pressures throughout five
orders of magnitude on a logarithmic scale, from 1~kbar to 10~Mbar. The simulations
were carried out in the following way. 1) A random arrangement of 256 \csixty{} units
was generated as in the previous section. 2) For each pressure value a 200~ps
equilibration was carried out at 1000~K using the Berendsen thermostat with the
same parameters
as in the previous section, and the Berendsen barostat with time constant of 1~ps
and bulk modulus of $\gamma_p = (\sqrt{P/1000} / 4.5 \times 10^{-5})$~bar, where
$P$ is in units of bar; in
the TurboGAP implementation, the bulk modulus is specified in units of the approximate
inverse compressibility of water, $1 / (4.5 \times 10^{-5}$~bar$^{-1})$. We find
that a variable
bulk modulus is necessary to accommodate the changing elastic response of the system
as it becomes compressed. 3) After the equilibration phase, the barostat is turned
off and a box scaling transformation is used to very slowly change the lattice vectors,
starting from the last snapshot of the equilibration and ending at the average lattice
vectors during the part of the equilibration period where the potential energy has settled. This
is done over 20~ps of MD. 4) A fixed-volume simulation is run for another 20~ps with only
the thermostat turned on, and structural data (atomic positions) is gathered for further
analysis. 5) The last snapshot from step 4) is taken as starting configuration for a
new step 2), increasing the temperature by 1000~K. Therefore, for each pressure there is
a continuous transformation in temperature from 1000~K to 5000~K, at 1000~K steps,
throughout all the stages detailed above.

The results of our simulations are summarized in \fig{04} and Table~\ref{03}. We find that
the region of metastability for a \csixty{} condensed phase lies in the lower left quadrant
of the metastable phase diagram at and below approximately 3000~K and 100~kbar. At higher temperatures
the \csixty{} system transforms into either liquid C (l-C) or graphite/graphitic C, depending
on the pressure, whereas
at higher pressures the transition is to extremely highly $sp^3$-rich amorphous C (a-C),
which we have denoted on the graph as amorphous diamond (a-D). At $T = 5000$~K and 1~Mbar we
start to observe crystalline diamond nucleation and grain formation, identified at ``polycrystalline''
diamond on the figure.
For very high pressures (10~Mbar),
the density (Table~\ref{03}) rises to more than twice the density of crystalline diamond at ambient
pressure ($\sim 3.5$~g/cm$^3$). This high-pressure phase shows, besides high $sp^3$ content in
the vicinity of 75~\%, also sizable exotic coordinations where C atoms can be surrounded by 5 and
more nearest neighbors. We have thoroughly studied 5-fold coordinated carbon complexes in the
context of GAP simulation in our previous work~\cite{caro_2020c}.
We have based the neighbor counts on a cutoff scheme, which is commonplace
in the literature and usually chosen at 1.85~\AA{}, the first minimum in the RDF at ambient
conditions. Due to the extreme compression of a-D in \fig{04}, we adjusted this cutoff to account
for the increase in mass density compared to crystalline diamond at ambient conditions,
$r_\text{cut} = 1.85~\text{\AA} \times \sqrt[3]{\rho_\text{diamond}/\rho}$.

The structure of the \csixty{} condensed phase was probed using the SOAP similarity metric
as in the previous section. We found that the structure is not fcc anymore at the probed
pressure/temperature combinations, even though we chose to simulate 256 \csixty{} units
precisely to accommodate a possible fcc arrangement. While the high-pressure \csixty{}
structure resembles hcp more than either fcc or bcc, its overall resemblance to hcp
decreases, compared to clusters at low
pressure. For this reason we have denoted the structure as hcp ``like'' in \fig{04}.

To elucidate the true nature of the liquid structures, in \fig{09} we show vibrational
density of states (DoS) calculations carried out with the DoSPT
code~\cite{caro_2016,caro_2017b,ref_dospt}. The vibrational DoS spectra show characteristic
profiles with strong self diffusion of the individual C atoms, which is proportional to the
zero-frequency DoS~\cite{mcquarrie_1976,lin_2003}. The self-diffusion constant
decreases as pressure increases. Together
with the high temperature structures identified as ``liquid C'' in \fig{04}, we have also
included what we identify as ``polycrystalline diamond''. The vibrational DoS for that structure
shows the bimodal features typical of diamond, and no self diffusion, highlighting that, even at
5000~K, 10~Mbar of pressure is enough to suppress liquid behavior from the sample. Finally, we
note the high-frequency peaks at low pressure, at around 50~THz, which correspond to the
$sp$ chains present in those liquid samples (and which are visible in \fig{04}).

\section{Code and potential availability}

The GAP and TurboGAP codes are freely available online for non-commercial academic
research~\cite{ref_quip,ref_turbogap}. The soap\_turbo library with modified SOAP
descriptors~\cite{caro_2019} is also freely available under the same terms. QUIP~\cite{ref_quip}
and the C60 GAP~\cite{muhli_2021} are freely available online, for non-commercial
as well as commercial research.

\section{Summary and outlook}

We have presented a complete methodology for accurate incorporation of vdW
interactions to ML interatomic potentials, based on the Tkatchenko-Scheffler
approach. The new approach has been implemented in the GAP and TurboGAP codes
and is freely available for academic research. Excellent computational performance
can be achieved, with only minor to moderate increase in CPU cost compared to
simulations without vdW corrections.
We have trained a new ML GAP potential for carbon enabled with the new methodology
and optimized it specifically for simulation of \csixty{}. The new force field has been
validated as an excellent general-purpose carbon potential and utilized to chart the phase
transformations taking place in \csixty{} under a wide range of temperatures and pressures,
reproducing many of the features observed experimentally.

We expect that the new C60 GAP potential will open the door for accurate simulation
of carbon materials and \csixty{} in particular. There is interest in \csixty{} as
precursor material for other carbon-based allotropes synthesized under extreme
conditions, and we believe that the C60 GAP may guide future experimental efforts in
this regard. Methodology wise, the next steps to build up and improve the inclusion
of vdW interactions from the current work will focus on
reproducing more accurate schemes like PBE+MBD~\cite{tkatchenko_2012} and
vdW-DF~\cite{dion_2004}, as well as fitting new vdW-inclusive
GAPs for materials other than carbon.
Extending
our methodology to other non-ionic materials is straightforward, and preliminary
work on vdW-inclusive general-purpose GAPs for systems containing C, H and O atoms
indicate that the approach works similarly well. On the other hand, systems where
significant charge transfer occurs (e.g., when ions are present) will require further
development to take into account the nonlocality of the effective Hirshfeld volumes.
If they are to remain computationally tractable, and thus amenable to large-scale MD
simulation, these new approaches may require a combination of short-range many-body
descriptors and simple (e.g., pairwise) long-range ones. All in all, we expect
rapid advances in the development of accurate and computationally efficient dispersion
correction schemes for ML-driven atomistic modeling in the near future.

\begin{acknowledgments}
The authors acknowledge funding from the Academy of Finland, under projects
310574, 330488 and 329483 (M.~A.~C.), 321713 (M.~A.~C, P. H.-L. and H.~M.),
308647 (X.~C.), 314298 (X.~C. and H.~M.), and the QTF Center of Excellence
program grant no. 312298 (T.A.-N.). M.~A.~C., P. H.-L. and H.~M.
also acknowledge a Seed Funding grant
from the Aalto University Materials Platform.
Computing time from CSC -- IT Center for Science for the MaCaNa project
and from Aalto University's Science IT project is gratefully acknowledged.
\end{acknowledgments}

\end{document}